\documentclass[10pt,conference]{IEEEtran}
\IEEEoverridecommandlockouts
\usepackage{cite}
\usepackage{amsmath,amssymb,amsfonts}
\usepackage{algorithmic}
\usepackage{graphicx}
\usepackage{textcomp}
\usepackage{xcolor}
\usepackage{url}
\usepackage{xurl}
\def\BibTeX{{\rm B\kern-.05em{\sc i\kern-.025em b}\kern-.08em
    T\kern-.1667em\lower.7ex\hbox{E}\kern-.125emX}}

\usepackage{xspace}
\usepackage[breakable]{tcolorbox}
\usepackage{soul}
\usepackage{xparse}
\usepackage{enumitem}
\usepackage{caption}
\usepackage{booktabs}
\usepackage{multirow}
\usepackage{tabularx}
\usepackage{array}
\usepackage{subcaption}
\usepackage{adjustbox}
\usepackage{listings}
\usepackage{balance}

\graphicspath{{figure/}}
\setlist[itemize]{leftmargin=1.5em}
\setlist[enumerate]{leftmargin=1.5em}

\newcommand{\toolname}{MoST\xspace}

\newcommand{\mainpoint}[1]{\par\smallskip\noindent\textbf{#1}}

\DeclareRobustCommand{\code}[1]{%
  {\small\ttfamily\begingroup\Urlmuskip=0mu plus 1mu\relax\path{#1}\endgroup}%
}

\newtcolorbox{rqbox}{breakable,left=2pt,right=2pt,top=2pt,bottom=2pt}

\title{Multi-Source and Cross-Scenario Strategy-Guided Code Optimization}
\author{
\IEEEauthorblockN{Yuwei Zhao\textsuperscript{1},
Qianyu Xiao\textsuperscript{1},
Ye Cui\textsuperscript{1},
Yijun Yu\textsuperscript{2}, and
Yingfei Xiong\textsuperscript{1,*}}
\IEEEauthorblockA{\textsuperscript{1}Key Laboratory of High Confidence Software Technologies
(Peking University), Ministry of Education;\\
School of Computer Science, Peking University, Beijing, China\\
\textsuperscript{2}The Open University, Milton Keynes, United Kingdom\\
\texttt{zhaoyuwei@stu.pku.edu.cn}, \texttt{norn773@163.com},
\texttt{yecui@pku.edu.cn}\\
\texttt{y.yu@open.ac.uk}, \texttt{xiongyf@pku.edu.cn}}
\thanks{\textsuperscript{*}Corresponding author: Yingfei Xiong
(\texttt{xiongyf@pku.edu.cn}).}
}

\begin{document}
\maketitle

\begin{abstract}

Automated code optimization improves program performance by refactoring source code, and recent studies use LLMs to generate optimization patches. The newest approaches are strategy-guided: they summarize strategies from historical optimization commits as static analysis rules, and use these rules to match code locations for LLMs to optimize. However, these approaches have two limitations: (1) the strategies may come from other \emph{knowledge sources}, such as textbooks and web pages, but the existing approaches cannot utilize them; (2) a strategy may be applicable to \emph{different scenarios}, e.g., different programming languages, but existing approaches can only formalize strategies for the scenario to which the source commit belongs.

To address these limitations, we propose \toolname, an LLM-based code optimization framework that integrates multiple knowledge sources across scenarios. \toolname uniformly represents items in different knowledge sources as \emph{evidence objects}, clusters them in a cross-source and cross-scenario manner to identify strategies, and transfers them to the target scenario when necessary for generating static analysis rules. To implement this process, \toolname employs a novel \emph{self-balanced} weighted clustering algorithm to balance evidence objects from different knowledge sources, and a novel \emph{example transfer} procedure to ensure the quality of the generated rules when transferring across scenarios.

On a benchmark containing 151 C/C++, 150 Python, and 50 Rust historical optimization tasks, compared with SemOpt, \toolname yields 24.44\%-180.00\% and 21.88\%-37.50\% more patches that are exactly the same as or semantically equivalent to developer patches, respectively. When optimizing 15 real-world projects, \toolname achieves 19.72\%-717.42\% maximum improvements and 4.44\%-258.17\% average improvements for the performance tests in the projects, significantly outperforming SemOpt and Codex.
\end{abstract}

\section{Introduction}\label{sec:introduction}\label{sec-intro}

\mainpoint{Background.}
Automated code optimization refactors source code to improve program performance. Inefficient code can increase execution time, resource consumption, and operational cost, and affect user experience and service quality~\cite{fur2011systems,nistor2013discovering,jovic2011catch,dean2014perfscope}. Although modern compilers implement small optimizations such as peephole optimization and common subexpression elimination~\cite{mckeeman1965peephole,cocke1970global}, many performance issues in real projects require large changes beyond what compilers can handle.


Large language models (LLMs) provide a new opportunity for code optimization. Code LLMs have shown capabilities in code understanding, generation, repair, and editing~\cite{nijkamp2022codegen,li2023starcoder,guo2024deepseek,fan2023large}, and recent approaches utilize LLMs to change source code to optimize program performance~\cite{shypula2023learning,garg2023rapgen,gao2024search,peng2025perfcodegen}. However, directly asking an LLM to optimize code remains unstable: the model may miss optimizable code locations, or ignore a useful optimization strategy at the target location.
As a result, the newest trend is \emph{strategy-guided}: mining useful optimization strategies from historical commits to guide LLMs in code optimization~\cite{semopt,garg2023rapgen}.

SemOpt is the state-of-the-art (SOTA) strategy-guided code optimization approach~\cite{semopt}. It extracts and clusters optimization strategies from historical optimization commits, and generates static analysis rules in Semgrep~\cite{semgrep} for each cluster. These rules are then used to scan target code and locate positions where the strategy may apply. SemOpt then provides the matched location, strategy description, and related examples to an LLM for optimization patch generation. 
In this way, SemOpt generates significantly better optimization patches than directly asking the LLM to optimize code.

\mainpoint{Problem and Challenges.}
Existing strategy-guided methods, including SemOpt and others, still have two limitations. First, the strategy library is built from one class of \emph{knowledge sources}: historical optimization commits. Optimization strategies may also appear in textbooks, manuals, Web pages, and other sources, but existing strategy-guided approaches cannot utilize them. Second, a strategy is formalized only for the \emph{scenario} of its source evidence. Here a scenario denotes the context in which a strategy applies, such as a programming language and a target CPU architecture. For example, a Semgrep rule generated from a Java commit cannot be used for C code, though it may describe a general optimization strategy for any imperative programming language.

\mainpoint{Our Approach.}
To address these limitations, we propose \textbf{\toolname}, a novel framework for \textbf{M}ulti-s\textbf{o}urce strategy construction and cross-\textbf{S}cenario strategy \textbf{T}ransfer in LLM-based automatic code optimization.
The basic idea is that LLMs have strong semantic understanding capabilities, and can identify and transfer optimization strategies across heterogeneous knowledge sources and scenarios.
To uniformly represent the data items containing strategies in heterogeneous knowledge sources, we introduce the concept of \emph{evidence object}, capturing the description, the example, and the applicable scenarios of the strategy. With the help of LLMs, the data items in heterogeneous knowledge sources are first converted uniformly into evidence objects, then clustered based on the description to identify strategies in a cross-source and cross-scenario manner, and finally used to generate static analysis rules in the target scenario.

Implementing this process has multiple challenges.
\begin{itemize}[leftmargin=*]
  \item \textbf{Balancing heterogeneous evidence during clustering (C1).} The evidence objects from different knowledge sources may have different frequencies and qualities. For example, the commits in existing projects may be more noisy than optimization manuals, and contain repetitive strategies more often. Equal-weight clustering can let frequent commit evidence dominate and discard less frequent but valuable manual evidence.

  To address this challenge, we introduce a novel self-balanced weighted clustering algorithm. This algorithm first searches for suitable weights for different knowledge sources plus other super parameters, and then applies weighted clustering such that different knowledge sources are balanced.

  \item \textbf{Reliable rule generation for cross-scenario transfer (C2).} When a strategy cluster lacks sufficient examples in the target scenario, \toolname has to construct the static analysis rules based on examples in other scenarios. We find that such a direct cross-scenario generation often leads to incorrect rules.

  To address this challenge, we introduce an example transfer procedure to first transfer examples from other scenarios to the target scenario. These examples are used not only to guide the generation of the rules, but also to validate the rules by checking whether a rule matches the pre-optimization example and does not match the post-optimization example.
\end{itemize}

\smallskip

\mainpoint{Evaluation.}
We evaluate \toolname at two levels. First, the historical optimization reproduction experiment in Section~\ref{sec:eval1} evaluates whether \toolname can reproduce historical patches written by human developers. The experiment covers 151 C/C++ tasks, 150 Python tasks, and 50 Rust tasks. Compared with SemOpt, \toolname yields 24.44\%--180.00\% and 21.88\%--37.50\% more patches that are exactly the same as or semantically equivalent to developer patches, respectively. Second, the real-world project optimization experiment in Section~\ref{sec:real-world-project-optimization} applies \toolname to 15 real-world projects. \toolname achieves 19.72\%--717.42\% maximum improvements and 4.44\%--258.17\% average improvements for the performance tests in these projects, significantly outperforming SemOpt and Codex.

\mainpoint{Contribution.}
This paper makes the following contributions:
\begin{itemize}
  \item We formulate multi-source, cross-scenario strategy reuse for LLM-based code optimization, and introduce evidence objects for uniformly representing data items from heterogeneous knowledge sources and for transferring strategies across scenarios (Section~\ref{sec:approach}).
  \item We propose self-balanced weighted clustering for balancing evidence objects from different knowledge sources, and an example transfer procedure for ensuring the quality of the generated static analysis rules for cross-scenario rule generation (Section~\ref{sec:approach}).
  \item We evaluate \toolname on 351 historical optimization tasks and 15 real-world projects. The results show that \toolname produces more developer-equivalent patches and achieves higher performance improvement on real-world projects than all baselines (Sections~\ref{sec:eval1} and~\ref{sec:real-world-project-optimization}).
\end{itemize}

\section{Motivating Example and Approach Overview}\label{sec:motivating-example}\label{sec:motivation}\label{sec-example}

\begin{figure*}[t]
  \centering
  \begingroup
  \lstset{
    basicstyle=\ttfamily\fontsize{6}{6.8}\selectfont,
    breaklines=true,
    columns=fullflexible,
    frame=single,
    literate={_}{{\textunderscore}}1,
    aboveskip=0pt,
    belowskip=0pt,
    xleftmargin=0pt,
    xrightmargin=0pt
  }
  \begin{subfigure}[t]{0.56\textwidth}
    \centering
    \vspace{0pt}
    \fbox{\begin{minipage}{0.94\linewidth}
      \scriptsize
      \textbf{5.5.3 Loop Blocking.}
      Loop blocking optimizes memory performance by reducing cache misses.  It
      transforms a problem's memory domain into smaller chunks instead of
      traversing the entire domain sequentially.  Each chunk is chosen to fit
      the data for a computation in cache, maximizing data reuse.  The manual
      also treats loop blocking as strip mining in two or more dimensions, and
      explains that blocking arrays into cache-sized rectangular chunks can
      eliminate redundant cache misses.
      \par\centering\ldots\par
      \raggedright
      {\ttfamily\fontsize{5}{5.4}\selectfont
      \begin{tabular}{@{}l@{}}
      A. Original Loop\\
      float A[MAX, MAX], B[MAX, MAX]\\
      for (i=0; i\textless MAX; i++) \{\\
      \quad for (j=0; j\textless MAX; j++) \{\\
      \quad\quad A[i,j] = A[i,j] + B[j, i];\\
      \quad \}\\
      \}\\
      \end{tabular}}
      \par\centering\ldots\par
      \raggedright
      {\ttfamily\fontsize{5}{5.4}\selectfont
      \begin{tabular}{@{}l@{}}
      B. Transformed Loop after Blocking\\
      float A[MAX, MAX], B[MAX, MAX];\\
      for (i=0; i\textless MAX; i+=block\textunderscore size) \{\\
      \quad for (j=0; j\textless MAX; j+=block\textunderscore size) \{\\
      \quad\quad for (ii=i; ii\textless i+block\textunderscore size; ii++) \{\\
      \quad\quad\quad for (jj=j; jj\textless j+block\textunderscore size; jj++) \{\\
      \quad\quad\quad\quad A[ii,jj] = A[ii,jj] + B[jj, ii];\\
      \quad\quad\quad \}\\
      \quad\quad \}\\
      \quad \}\\
      \}\\
      \end{tabular}}
    \end{minipage}}

    \caption{Original document excerpt with strategy text and partial examples}
    \label{fig:motivation-doc-strategy}
    \label{fig:motivation-doc-fragment}
    \label{fig:motivation-doc-before}
    \label{fig:motivation-doc-after}
  \end{subfigure}
  \hfill
  \begin{minipage}[t]{0.41\textwidth}
    \vspace{0.3em}
  \begin{subfigure}[t]{\linewidth}
    \centering
    \vspace{0pt}
\begin{lstlisting}
fn blur(src: &[f32], dst: &mut [f32], w: usize, h: usize) {
    ...
    for y in 1..h - 1 {
        for x in 1..w - 1 {
            dst[y * w + x] = stencil(src, w, y, x);
        }
        ...
\end{lstlisting}
    \caption{Before-optimization code in the Rust target}
    \label{fig:motivation-rust-before}
  \end{subfigure}
  \par\vspace{0.45em}
  \begin{subfigure}[t]{\linewidth}
    \centering
    \vspace{0pt}
\begin{lstlisting}
fn blur(src: &[f32], dst: &mut [f32], w: usize, h: usize) {
    ...
    const TILE: usize = 64;
    for y0 in (1..h - 1).step_by(TILE) {
        for x0 in (1..w - 1).step_by(TILE) {
            for y in y0..(y0 + TILE).min(h - 1) {
                for x in x0..(x0 + TILE).min(w - 1) {
                    dst[y * w + x] = stencil(src, w, y, x);
                }
                ...
\end{lstlisting}
    \caption{After-optimization code in the Rust target}
    \label{fig:motivation-rust-after}
  \end{subfigure}
  \end{minipage}
  \endgroup
  \caption{A document-sourced, cross-scenario motivating example.  The Intel
  optimization document provides excerpted strategy prose and partial C examples rather
  than a commit-style before/after pair.  The Rust target applies the same
  traversal strategy in another language.}
  \label{fig:motivation-example}
\end{figure*}

\subsection{An Optimization Strategy in a Document}
\label{sec:motivation-concrete-opportunity}

Figure~\ref{fig:motivation-doc-strategy} reproduces an excerpt from the Intel
optimization document~\cite{intel64ia32optimization}, with less relevant text
omitted.  The excerpt interleaves prose and partial C examples.  Its prose
describes loop blocking as dividing large traversals into cache-friendly blocks
to reduce cache misses and improve data reuse, and its example rewrites a
two-dimensional traversal over \code{A} and \code{B} by advancing outer indices
by \code{block_size}.

Figures~\ref{fig:motivation-rust-before}
and~\ref{fig:motivation-rust-after} show the corresponding opportunity in a
Rust loop.  The original loop visits interior index pairs row by row, while the
optimized loop visits \code{TILE}-sized regions to improve cache efficiency.

At a high level, both changes replace full traversal with cache-friendly blocked
traversal, so the strategy from the document can be used to optimize the Rust code.

\subsection{SemOpt and its Limitations}
\label{sec:motivation-limitations}

As introduced before, SemOpt~\cite{semopt} is the SOTA strategy-guided code optimization approach.  Given a corpus
of historical optimization commits, SemOpt first generates strategy descriptions from the commits and clusters the commits by the descriptions.  In this way,
each cluster represents an optimization strategy.  SemOpt then generates static
analysis rules from the commits in this cluster, and uses the rules to scan
target code for optimization opportunities.  When a rule matches, SemOpt asks
an LLM to generate a patch by prompting it with the matched code, the strategy
description, and the examples from the cluster.  In this way, SemOpt guides the
LLM to generate significantly better optimization patches with the precise
location and strategy information.

The running example demonstrates the limitations of SemOpt and other strategy-guided approaches. First, the source data is the raw document excerpt in Figure~\ref{fig:motivation-doc-strategy}, where prose and partial examples are mixed, while existing strategy-guided approaches could only utilize commits but not other \emph{knowledge sources}. As a result, SemOpt cannot extract and normalize this document information, and the loop-blocking strategy would not enter its strategy library. Second, the strategy is expressed with C snippets and partial examples such that existing approaches could only generate static analysis rules for C programs based on the examples, while the target program in Figures~\ref{fig:motivation-rust-before} and~\ref{fig:motivation-rust-after} is in Rust, a different application \emph{scenario}.



\subsection{Approach Overview} 
\label{sec:motivation-solution}

\begin{figure*}[t]
    \centering
    \includegraphics[width=\linewidth]{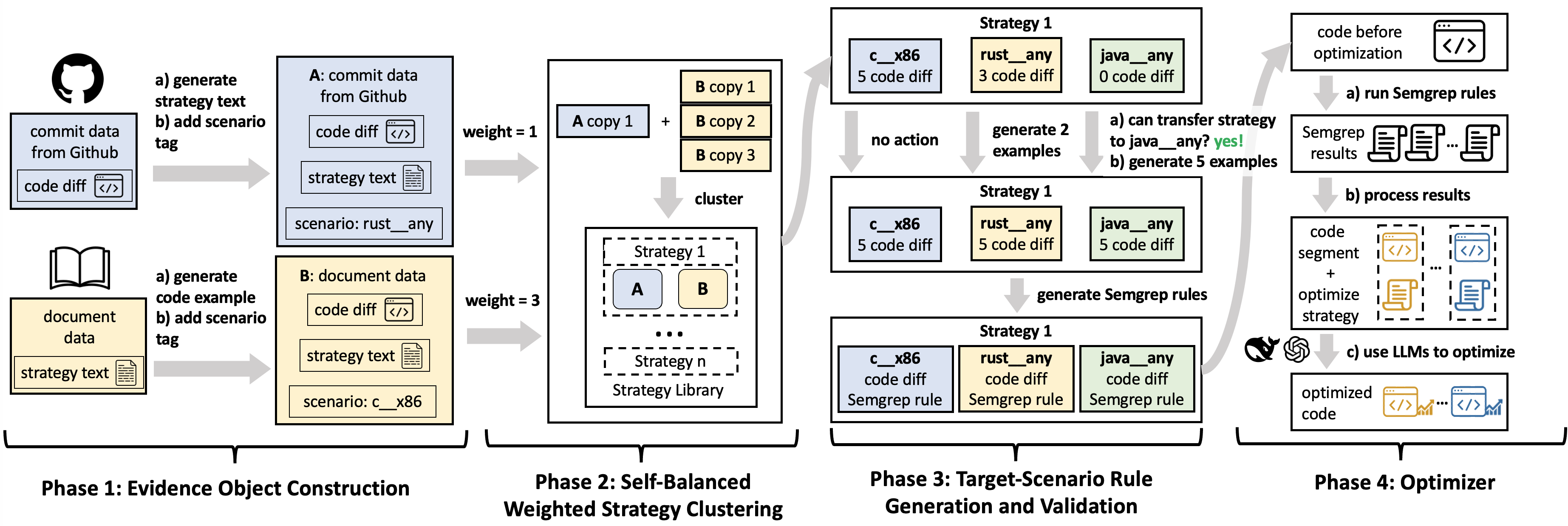}
    \caption{The overall workflow of \toolname. It first constructs evidence
    objects from multiple sources, then discovers strategy clusters, generates
    target-scenario rules through example transfer and validation, and finally
    uses validated rules to guide LLM-based code optimization.}
    \label{fig:approach-overview}
\end{figure*}

Figure~\ref{fig:approach-overview} summarizes how \toolname addresses the
example.

First, we notice that though strategies may exist in heterogeneous data items in different knowledge sources, these data items can all be represented by several basic components: (1) A natural language (NL) description about when and how to apply the strategy, capturing the nature of the strategy beyond specific application scenarios and supporting cross-scenario clustering. In our example, Figure~\ref{fig:motivation-doc-strategy} is the raw PDF excerpt, from whose prose \toolname extracts the NL description of the C loop-blocking strategy. (2) A code-modification example consisting of a before-optimization code snippet and an after-optimization code snippet, useful in guiding the LLM to generate static analysis rules. The same excerpt contains partial C examples, so \toolname normalizes them into complete before- and after-optimization snippets when constructing this field. (3) A scenario tag denoting the scenario of the code-modification example. For example, \code{c__x86_64} describes a C-language optimization for the \code{x86_64} architecture. We also allow a tag to capture multiple concrete scenarios, e.g., \code{c__any} describes a C-language optimization that can be applied to any architecture. We always use the most general scenario tag: the scenario for our example is \code{c__any}, as all popular architectures have cache. (4) A source type (e.g., commits, documents) records the type of the knowledge source where the data item comes from.

We call such a uniformly represented data item an \emph{evidence object}, and convert data items from different knowledge sources into evidence objects with the assistance of LLMs, corresponding to phase 1 in Figure~\ref{fig:approach-overview}. For example, given a document, we first divide it into multiple segments to fit into the context window, ask an LLM to identify evidence objects from each segment, and complete missing fields such as normalized before- and after-optimization snippets when necessary. Given a commit, we extract the code modification example directly from the commit, and ask an LLM to generate the NL description based on the example and the commit message.


Second, after we have a set of evidence objects, we can use the NL descriptions for clustering as in SemOpt. However, evidence objects from different knowledge sources differ in frequency and quality. For example, evidence objects from documents are usually of high quality and rarely duplicated, while evidence objects from commits often contain noise and duplicated strategies. If we directly cluster the two, the evidence objects from the documents would be treated as noise and ignored, though they in fact have better quality.

To address this issue, in phase 2, \toolname uses weighted density clustering~\cite{ester1996density}, so document evidence can receive a higher weight and avoid being treated as noise. This requires selecting the document weight, description-similarity threshold, and minimum cluster size together, while the commit weight remains fixed at $1$.
\toolname employs the self-balanced selection criterion. The key insight is that a good parameter configuration should preserve the scenario-level strategy counts obtained from single-scenario clustering after all evidence objects are clustered together. The system samples a bounded set of parameter values and selects the one that minimizes this scenario-level deviation without rule generation or LLM optimization.


Third, in the original SemOpt, the next step is to sample commits from the cluster, generate $b$ Semgrep rules from each sampled commit, and use rule agreement to rank matched code locations. This process can improve optimization stability. However, in \toolname, a cluster may contain evidence from multiple scenarios, and directly generating a Rust rule from a C example can noticeably reduce rule correctness.

To address this issue, in phase 3, \toolname first computes a cluster weight $k$ and uses it as the example budget. The sampling pool follows source weights: an evidence object with integer weight $x$ contributes $x$ copies of its code-modification example, while \toolname prefers distinct examples and uses repeated copies only when distinct examples are insufficient. Target-scenario example construction then has three cases. If the cluster has at least $k$ target-scenario example copies, \toolname samples $k$ of them. If it has some but fewer than $k$, \toolname keeps them and instantiates the remaining examples from other-scenario references. If it has none, the strategy has not appeared in the target scenario before, so \toolname first checks whether the strategy applies to that scenario; only applicable strategies are instantiated from other-scenario references. After the example budget is filled, \toolname generates $b$ Semgrep rules from each target-scenario example. In our running example, \toolname first checks that loop blocking applies to Rust, then transfers the normalized C code-modification example to Rust.

To further ensure the quality of the generated rules, \toolname also adds functional validation after rule generation. For each rule generated from an example, we check whether the rule matches the before-optimization code and rejects the after-optimization code. If not, we ask the model to regenerate. Combining the two procedures, \toolname can generate high-quality rules from the evidence objects.


Finally, in phase 4, \toolname proceeds in the same way as SemOpt: scanning the code with the generated rules and prompting the LLM to optimize the most matched code locations with their respective strategies. In our running example, we would get the optimized Rust code after this step.


\section{Approach Details and Implementation}\label{sec:approach}\label{sec-approach}

This section details the four phases outlined in
Figure~\ref{fig:approach-overview}.

\subsection{Phase 1: Evidence object construction}\label{sec:approach-phase1}

Following the representation introduced in
Section~\ref{sec:motivation-solution}, \toolname maps each input item to an
evidence object
\begin{equation}
e=\langle P,E,\mathcal{T},y\rangle,
\label{eq:evidence-object}
\end{equation}
where $P$ is the NL description, $E$ is the before/after code-modification
example, $\mathcal{T}$ contains the applicable scenario tags, and $y$ is the
source type.  The source type remains attached so that later phases can trace
the object's origin and apply the corresponding clustering weight.

The conversion differs by knowledge-source type.  For commits, \toolname
follows the optimization-commit extraction process of SemOpt~\cite{semopt}: it
takes the code change and its context as $E$ and asks an LLM to summarize the
strategy as $P$.  For documents, the current implementation processes the
\emph{Intel\textregistered{} 64 and IA-32 Architectures Optimization Reference
Manual Volume 1}~\cite{intel64ia32optimization}.  It divides the extracted PDF
text into chunks, asks an LLM to identify optimization descriptions and
applicability constraints, and generates a before/after example when the
document does not provide one.  This process yields 48,440 commit-derived
evidence objects (27,463 C, 10,073 Python, 7,098 Java, and 3,806 Rust) and 189
document-derived evidence objects.

To assign $\mathcal{T}$, let $\mathcal{L}$ contain the properties used to
describe a scenario; the current implementation uses programming language and
target architecture.  A scenario tag pairs each property $l_i$ with a target
value $t_i$:
\begin{equation}
T=\bigl((l_i=t_i)\bigr)_{i=1}^{k},\qquad l_i\in\mathcal{L}.
\label{eq:scenario}
\end{equation}
A target is either one concrete value, such as \code{rust} or \code{x86_64},
or a named group of concrete values.  For example, \code{any} denotes the
group of all architectures considered by \toolname.  The system assigns the
most general valid target: an architecture-independent strategy receives
\code{any}, whereas a strategy that depends on an instruction set or
microarchitecture receives the corresponding concrete architecture.  A
target scenario matches a tag when its value in every property either equals
the tag's concrete value or belongs to the tag's group.

\subsection{Phase 2: Self-balanced weighted strategy clustering}\label{sec:approach-phase2}

As described in Section~\ref{sec:motivation-solution}, \toolname measures
similarity using only the NL descriptions and applies weighted density
clustering~\cite{ester1996density}.  The three clustering parameters are the
document weight $w_d$, the description-similarity threshold $\tau$, and the
minimum cluster size $s_{\min}$.  The commit weight is fixed at $1$, and
$s_{\min}$ specifies the minimum weighted evidence support required to retain a
cluster.

The self-balanced selection criterion chooses the three parameters jointly from
a bounded candidate set $\Theta$, sampled before evaluation as in random
hyperparameter search~\cite{bergstra2012random}.  For each candidate $\theta$
and scenario tag $T$, \toolname takes the single-scenario clustering result as
the reference count $K_T$, clusters all evidence objects using $\theta$, and
obtains $M_T(\theta)$, the retained native strategy-cluster count for $T$.  To
preserve this count without excessive inflation, and
following log-ratio analysis for relative data~\cite{aitchison1982statistical},
\toolname measures
\begin{equation}
r_T(\theta)=\log\frac{M_T(\theta)+\epsilon}{K_T+\epsilon},
\label{eq:balance-ratio}
\end{equation}
where $\epsilon>0$ avoids zero-count divisions. The per-group loss is
\begin{equation}
\ell_T(\theta)=
\left|r_T(\theta)\right|+\lambda\max(0,-r_T(\theta)).
\label{eq:balance-loss}
\end{equation}
The first term penalizes both under- and over-clustering, while the second adds
a larger penalty for losing strategy coverage.  For
$\mathcal{T}_+=\{T\mid K_T>0\}$, the criterion scores each candidate by
\begin{equation}
L(\theta)=\frac{1}{|\mathcal{T}_+|}
\sum_{T\in\mathcal{T}_+}\ell_T(\theta),
\label{eq:balance-objective}
\end{equation}
and selects
\begin{equation}
\theta^*=\arg\min_{\theta\in\Theta}L(\theta).
\label{eq:parameter-selection}
\end{equation}
This selection reruns only clustering, without LLM calls, Semgrep rule
generation, or performance evaluation.  The selected configuration is
document weight $3$, similarity threshold $0.76$, and minimum cluster size $5$;
Section~\ref{sec:eval1-rq2} compares this automatically selected configuration
with nearby alternatives.
After the selected clustering run, \toolname assigns each strategy cluster an
integer example budget:
\begin{equation}
W(C)=\left\lfloor
\frac{1}{|\mathcal{S}(C)|}\sum_{e\in\mathcal{U}(C)} w_{y(e)}
+\frac{1}{2}
\right\rfloor .
\label{eq:cluster-weight}
\end{equation}
Here, $\mathcal{U}(C)$ is the deduplicated evidence in $C$,
$\mathcal{S}(C)$ is the set of source scenarios represented in $C$, and
$w_{y(e)}$ is the selected source-type weight of evidence $e$.  The averaging
prevents a multi-scenario cluster from receiving a larger budget only because it
appears in more scenarios; $W(C)$ is used as the target-scenario example budget
in Phase~3.

\subsection{Phase 3: Target-scenario rule generation and validation}\label{sec:approach-phase3}

For each strategy cluster $C$ and target scenario $T^*$, \toolname sets
$n_C=W(C)$ and selects $n_C$ NL-description--example pairs from which to
generate rules.  For sampling, an evidence object with integer source weight
$x$ contributes $x$ copies of its example; \toolname prefers distinct examples
and uses repeated copies only when the distinct examples are insufficient.  Let
$m(C,T^*)$ be the number of weighted example copies in $C$ whose tags exactly
match $T^*$.  Since $n_C$ is rounded from the average weighted evidence
strength, the cluster-level sampling pool contains enough copies to fill the
budget from other scenarios when exact-target examples are missing.  The
selection has three cases:
\begin{enumerate}[leftmargin=*]
  \item If $m(C,T^*)\geq n_C$, \toolname samples $n_C$ exact-tag pairs.
  \item If $0<m(C,T^*)<n_C$, it retains the exact-tag pairs, samples the remaining
  $n_C-m(C,T^*)$ pairs from other scenarios in $C$, and asks an LLM to transfer
  their descriptions and examples to $T^*$.
  \item If $m(C,T^*)=0$, an LLM first determines whether the strategy applies
  to $T^*$.  If it does, \toolname transfers $n_C$ sampled pairs as above;
  otherwise, it skips $C$ for this target.
\end{enumerate}
Rule generation starts only after this step fills the target-scenario example
budget for the retained cluster--target pair.

For each selected or transferred pair, the LLM independently generates $b$
Semgrep rules~\cite{semgrep}.  Each rule is first checked for successful
parsing and execution.  It must then match the before-optimization code in the
example and reject the after-optimization code.  On a parsing error, false
negative, or false positive, \toolname returns the failure and relevant
location to the LLM for regeneration.  Only rules passing both checks are used
for $T^*$.  This check is stronger than SemOpt's execution-only test for LLM-generated
rules, which can admit rules with false positives, false negatives, or
unclear semantic boundaries~\cite{yang2025knighter,liu2024write,li2025automated}.

The selected clustering run produces 356 strategy clusters, including 39
cross-scenario clusters.  Table~\ref{tab:approach-scale} reports how these
clusters are used for each target scenario.  A cluster with \emph{exact-tag
evidence} contains at least one evidence object whose scenario tag is exactly
the target tag.  A cluster \emph{added from other-tag evidence} has no such
object, but passes the applicability check and supplies examples that are
transferred to the target scenario.  The third row sums these two groups, and
the final row reports the Semgrep rules generated from them.

\begin{table}[t]
\centering
\scriptsize
\setlength{\tabcolsep}{1.5pt}
\renewcommand{\arraystretch}{1.08}
\caption{Per-scenario strategy-cluster and rule-generation scale.}
\label{tab:approach-scale}
\begin{adjustbox}{max width=\columnwidth}
\newcommand{\tagcell}[2]{\begin{tabular}[c]{@{}c@{}}\textbf{#1\_\_}\\\textbf{#2}\end{tabular}}
\begin{tabular}{@{}>{\raggedright\arraybackslash}p{0.23\columnwidth}rrrrrrrr@{}}
\toprule
\textbf{Metric} & \tagcell{c}{any} & \tagcell{c}{arm64} & \tagcell{c}{x86\_64} & \tagcell{java}{any} & \tagcell{python}{any} & \tagcell{rust}{any} & \tagcell{rust}{arm64} & \tagcell{rust}{x86\_64} \\
\midrule
\textbf{Clusters with exact-tag evidence} & 271 & 0 & 5 & 39 & 63 & 38 & 0 & 0 \\
\textbf{Clusters added from other-tag evidence} & 24 & 24 & 24 & 104 & 85 & 130 & 133 & 137 \\
\textbf{All target-scenario clusters} & 295 & 24 & 29 & 143 & 148 & 168 & 133 & 137 \\
\textbf{Generated Semgrep rules} & 9,735 & 730 & 900 & 4,395 & 4,475 & 5,135 & 4,345 & 4,455 \\
\bottomrule
\end{tabular}
\end{adjustbox}
\end{table}

\subsection{Phase 4: Optimizer}\label{sec:approach-phase4}

The final phase follows SemOpt's optimizer~\cite{semopt}.  Given a target
function and scenario $T^*$, \toolname loads the validated rules whose tags
match $T^*$ and scans the function.  Each match identifies a code location and
an optimization strategy.  \toolname merges matches only when both are the
same, ranks each merged location--strategy pair by the number of supporting
rules, and retains the top $N=25$ pairs.

For each retained pair, the LLM receives the target function, matched location,
strategy description, related target-scenario examples, and context indicated
by $T^*$.  For an architecture-specific target, this context can include the
processor model, architecture, and available instruction sets.  The LLM then
generates the optimization patch.

\section{Evaluation 1 Historical Optimization Task Reproduction}\label{sec:eval1}

\begin{enumerate}[left=1.5em]
    \item[\textbf{RQ1}] Compared with the baselines, can \toolname produce more successful optimizations?
    \item[\textbf{RQ2}] How important are the two core design components of \toolname?
    \item[\textbf{RQ3}] Are the optimization suggestions generated by \toolname practically applicable?
\end{enumerate}

\subsection{Experimental Setup}\label{sec:eval1-setup}

\subsubsection{Benchmark}\label{sec:eval1-benchmark}

This experiment uses historical optimization tasks. Each task records the codebase, commit hash, and complete pre- and post-commit functions. The pre-commit function is the tool input; the post-commit function forms the developer patch and ground truth.

We directly reuse the SemOpt C/C++ and Python benchmarks, which contain 151 and 150 optimization tasks, respectively~\cite{semopt}. Because SemOpt does not include a Rust benchmark, we follow the same process to build one. From the 100 most-starred Rust codebases on GitHub, we collect 1,008,176 main-branch commits, retain commits modifying only a single Rust function, filter optimization-related commits by keywords and code modifications, obtain 732 commits, and randomly sample 50 optimization tasks.

Because the benchmarks come from historical GitHub commits, they may overlap with data sources used by \toolname and the baselines. To mitigate data leakage, we exclude exact matches to the current task commit or task code from all available data; the remaining data can still support strategy construction, retrieval, and optimization.

\subsubsection{Baselines}\label{sec:eval1-baselines}

We compare \toolname with three baselines.

\mainpoint{Direct Prompt}
This baseline provides the current function in a static prompt and asks the LLM to generate optimized code.

\mainpoint{Retrieval Augmented Generation (RAG)}
Following Gao et al.~\cite{gao2023makes}, RAG retrieves the four most similar optimization examples from the available data with BM25, sorts them in ascending similarity order, and prompts the LLM with them and the current code. The current task commit and exact code matches are excluded.

\mainpoint{SemOpt}
For C/C++ and Python, we reuse the original SemOpt tool implementation. Since the original SemOpt paper does not include a Rust version, we reimplement it according to the SemOpt workflow~\cite{semopt}.

We omit RAPGen~\cite{garg2023rapgen} and Clang-Tidy~\cite{clangtidy} because SemOpt has already systematically compared with them and shown that they are weaker on historical optimization tasks~\cite{semopt}.

\subsubsection{Evaluated LLMs}\label{sec:eval1-llms}

We use DeepSeek-V4-Pro~\cite{deepseek2026v4pro} in all experiments and additionally run GPT-5.2~\cite{openai2026gpt52} on C/C++. DeepSeek-V4-Pro is the primary optimizer model because it was strong and stable during our experiments. GPT-5.2 provides a supplementary closed-source setting for C/C++; due to budget constraints, we do not extend it to Python and Rust. We set temperature to 0 and keep all other provider settings at their defaults.

\subsubsection{Metrics}\label{sec:eval1-metrics}

To evaluate optimized-code correctness, we reuse the two metrics used by SemOpt~\cite{semopt}.
\begin{itemize}[leftmargin=*]
    \item \textbf{Exact Match (EM)} removes comments, indentation, and whitespace from the developer ground truth and generated code, and checks whether the normalized strings are identical.
    \item \textbf{Semantic Equivalence (SemEqv)} compares the generated patch with the developer patch and judges whether they are semantically equivalent.
\end{itemize}

\subsubsection{Implementation Details}\label{sec:eval1-implementation-details}

Following existing work, all LLMs use temperature 0~\cite{gao2023makes,semopt}. Each optimization task is repeated 3 times, and a method succeeds if at least one generation satisfies the corresponding metric. Following SemOpt's experimental setting~\cite{semopt}, we set the rule-generation budget to $b=5$ Semgrep rules per sampled commit or target-scenario example.

\subsection{RQ1 Comparison with Baselines}\label{sec:eval1-rq1}

To evaluate \toolname on historical optimization tasks, we compare it with Direct Prompt, RAG, and SemOpt. Figure~\ref{fig:eval1-rq1-results} reports successful-task proportions on the C/C++, Python, and Rust benchmarks, which contain 151, 150, and 50 tasks, respectively.

\begin{center}
\centering
\includegraphics[width=\linewidth]{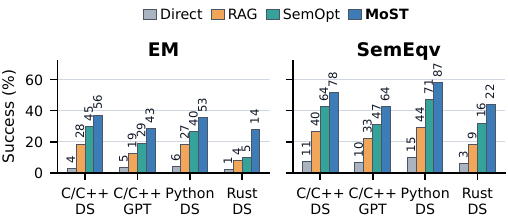}
\captionof{figure}{Performance comparison across languages and models. Bar heights show successful optimization proportions, and integer labels show successful-task counts. DS and GPT denote DeepSeek-V4-Pro and GPT-5.2, respectively.}
\label{fig:eval1-rq1-results}
\label{tab:eval1-rq1-results}
\end{center}

\subsubsection{Comparison to SemOpt}

As shown in Figure~\ref{fig:eval1-rq1-results}, \toolname obtains higher results than SemOpt under all language and model settings, improving EM by 24.44\%--180.00\% and SemEqv by 21.88\%--37.50\%. SemOpt derives optimization strategies from historical optimization commits and uses rules to locate optimizable code regions~\cite{semopt}, whereas \toolname uses multi-source evidence objects, transfers examples to target scenarios when needed, and validates target-scenario rules before optimization. These results suggest that multi-source strategy construction and reliable rule generation complement the coverage limits of a single historical-commit strategy library.

\subsubsection{Comparison to Direct Prompt and RAG}

Direct Prompt asks the LLM to generate an optimization result from the input function, mainly relying on the model's internal code knowledge and the prompt. RAG further provides similar historical optimization examples from the available data. However, neither method explicitly abstracts reusable optimization strategies or uses rules to locate where an optimization should be applied.

Figure~\ref{fig:eval1-rq1-results} also shows that \toolname outperforms Direct Prompt and RAG under all settings, improving successful optimizations by 480.00\%--1300.00\% and 93.94\%--250.00\%, respectively. Direct LLM generation is unstable for reproducing developer optimizations. RAG improves over Direct Prompt with similar examples, but remains below SemOpt and \toolname, suggesting that retrieved examples cannot replace strategy-level abstraction and rule-guided localization.

\subsubsection{Cross-language Results}

\toolname achieves the highest success rate on C/C++, Python, and Rust, so its benefit is not limited to one language. Compared with SemOpt, \toolname improves successful optimizations on Rust by 37.50\%--180.00\%, above the 21.88\%--48.28\% range on C/C++ and Python. Since C/C++ and Python have more historical optimization data, SemOpt's single-language strategy libraries already have relatively high coverage; Rust has less data and benefits more from multi-source strategy construction and cross-scenario transfer.

\begin{rqbox}
\textbf{Answer to RQ1}
\toolname produces more successful optimizations than Direct Prompt, RAG, and SemOpt on C/C++, Python, and Rust historical optimization tasks, improving over SemOpt by 24.44\%--180.00\% in EM and 21.88\%--37.50\% in SemEqv, and improving the number of successful optimizations over Direct Prompt and RAG by 480.00\%--1300.00\% and 93.94\%--250.00\%, respectively.
\end{rqbox}

\subsection{RQ2 Ablation Study}\label{sec:eval1-rq2}

We evaluate the two core components of \toolname on the C/C++ benchmark with DeepSeek-V4-Pro fixed as the evaluated LLM. Section~\ref{sec:approach-phase2} describes how \toolname selects document weight $3$, similarity threshold $0.76$, and minimum cluster size $5$ by minimizing a clustering-only balance loss.
The loss uses scenario-level strategy-count deviation with $\epsilon=1$ and $\lambda=1$, and requires no rule generation, LLM optimization, or performance evaluation. To check whether this low-cost selection matches optimization outcomes, we use a disjoint 45-task pilot set, with 15 tasks per language and no overlap with the formal benchmark; the C/C++ and Python sets reuse the SemOpt setting~\cite{semopt}, and the Rust set samples from GitHub-star ranks 101--200 after the same filtering process.

\begin{table}[t]
\centering
\scriptsize
\setlength{\tabcolsep}{2.4pt}
\caption{Strategy-library parameter configurations.}
\label{tab:eval1-parameter-selection}
\label{tab:eval1-pilot}
\begin{adjustbox}{max width=\linewidth}
\begin{tabular}{@{}lcccccccccc@{}}
\toprule
\textbf{Param.} & \textbf{C1} & \textbf{C2} & \textbf{C3} & \textbf{C4} & \textbf{C5} & \textbf{C6} & \textbf{C7} & \textbf{C8} & \textbf{C9} & \textbf{C10} \\
\midrule
Doc. weight & 3 & 3 & \textbf{3} & 3 & 4 & 4 & 5 & 5 & 3 & 5 \\
Threshold & 0.78 & 0.77 & \textbf{0.76} & 0.74 & 0.90 & 0.91 & 0.90 & 0.91 & 0.91 & 0.84 \\
Min size & 5 & 5 & \textbf{5} & 6 & 4 & 4 & 5 & 5 & 3 & 5 \\
Balance loss $\downarrow$ & 3.57 & 3.26 & \textbf{2.93} & 3.09 & 4.58 & 4.72 & 5.48 & 5.68 & 3.44 & 4.01 \\
EM & 12 & 13 & \textbf{14} & 11 & 12 & 12 & 9 & 9 & 13 & 10 \\
\bottomrule
\end{tabular}
\end{adjustbox}
\end{table}

Table~\ref{tab:eval1-parameter-selection} reports the ten parameter configurations, their clustering-only balance losses, and pilot-study EM results. C3 has both the lowest loss and the best pilot-study EM, supporting the parameter-selection criterion. Table~\ref{tab:eval1-rq2} reports the complete method and two ablation variants.

\begin{table}[t]
\centering
\scriptsize
\renewcommand{\arraystretch}{1.05}
\caption{Ablation study on the C/C++ benchmark with DeepSeek-V4-Pro.}
\label{tab:eval1-rq2}
\begin{adjustbox}{max width=\linewidth}
\begin{tabular}{@{}lcc@{}}
\toprule
\textbf{Approach} & \textbf{EM} & \textbf{SemEqv} \\
\midrule
\toolname & \textbf{56 (37.09\%)} & \textbf{78 (51.66\%)} \\
w/o weighted clustering & 50 (33.11\%) & 71 (47.02\%) \\
w/o reliable rule generation & 28 (18.54\%) & 44 (29.14\%) \\
\bottomrule
\end{tabular}
\end{adjustbox}
\end{table}

\mainpoint{w/o weighted clustering.}
This variant sets all source-type weights to $1$ and keeps other settings unchanged, evaluating source weights and weighted clustering.

\mainpoint{w/o reliable rule generation.}
This variant disables example transfer and directly asks the LLM to generate Semgrep rules from available cluster evidence. It also disables functional validation, so rules are not checked against target-scenario before/after examples.

As shown in Table~\ref{tab:eval1-rq2}, removing weighted clustering reduces successful optimizations by 8.97\%--10.71\%, and removing reliable rule generation causes a larger 43.59\%--50.00\% decrease. These results suggest that weighting helps protect high-value but low-frequency sources, while target-scenario examples and functional validation are important for reliable rule-based localization.

\begin{rqbox}
\textbf{Answer to RQ2}
The balance loss selects the configuration with the best pilot-study EM; removing weighted clustering and reliable rule generation reduces successful optimizations by 8.97\%--10.71\% and 43.59\%--50.00\%, respectively.
\end{rqbox}

\subsection{RQ3 Practical Applicability}\label{sec:eval1-rq3}

To evaluate \toolname's practical value, we measure how many optimization suggestions are acceptable in realistic settings. We consider a suggestion valuable if it preserves the original semantics and improves performance, such as execution speed or resource efficiency, even without exactly matching the original commit's optimization.

Due to manual-inspection cost, we evaluate this question only on the C/C++ benchmark with DeepSeek-V4-Pro. Among \toolname's optimization suggestions, \textbf{89.70\%} satisfy this criterion, indicating that most are practically applicable in realistic settings. This rate is close to the 89.86\% reported by SemOpt with DeepSeek-V3~\cite{semopt}, suggesting comparable practical applicability.

\begin{rqbox}
\textbf{Answer to RQ3}
On the C/C++ benchmark with DeepSeek-V4-Pro, \textbf{89.70\%} of \toolname's optimization suggestions are manually judged acceptable, close to SemOpt's reported 89.86\% with DeepSeek-V3.
\end{rqbox}

\section{Real-world Project Optimization}\label{sec:real-world-project-optimization}

\begin{enumerate}[left=1.5em]
    \item[\textbf{RQ4:}] On real-world projects, can \toolname produce more correct optimization results with performance improvements than SemOpt and Codex Agent?
    \item[\textbf{RQ5:}] What source and transfer provenance do \toolname's effective optimization results have?
\end{enumerate}

\subsection{Experimental Setup}\label{sec:real-world-setup}

\subsubsection{Benchmark Projects}\label{sec:real-world-benchmark}

We evaluate \toolname on five C/C++ projects, five Python projects, and five Rust projects. The ten C/C++ and Python projects reuse the real-world project optimization subjects from SemOpt~\cite{semopt}. The five Rust projects are newly collected to evaluate practical optimization after extending cross-scenario transfer to Rust.

\textbf{C/C++ Projects:} We reuse the five C/C++ projects from SemOpt: RocksDB~\cite{rocksdb}, Redis~\cite{redis}, gRPC~\cite{grpc}, LevelDB~\cite{leveldb}, and spdlog~\cite{spdlog}.

\textbf{Python Projects:} We reuse the five Python projects from SemOpt: Click~\cite{click}, Flask~\cite{flask}, Jinja2~\cite{jinja}, Requests~\cite{requests}, and Scrapy~\cite{scrapy}.

\textbf{Rust Projects:} We collect Rapier~\cite{rapier}, image~\cite{rustimage}, regex~\cite{rustregex}, sqlparser~\cite{sqlparser}, and Tokio~\cite{tokio}. These projects cover a physics engine, image processing, regular expressions, SQL parsing, and asynchronous runtime support, representing different types of performance-sensitive Rust code. We select them for stable buildable versions, runnable performance tests, and performance-critical code suitable for automated optimization.

Each project provides a relatively complete performance test suite for quantitative optimization measurement. For each project, we clone a fixed version, write scripts to compile C/C++ and Rust code or install the Python package, and run built-in unit and performance tests. The 15 projects contain 259 performance data points in total, with 2--45 per project. Each data point usually reflects one functionality or performance metric.

\subsubsection{Baselines}\label{sec:real-world-baselines}

We compare two types of baselines and report \toolname as the method proposed in this paper.

\mainpoint{SemOpt:}
SemOpt is the original strategy-guided code optimization method. We follow and reuse the implementation and workflow from its paper~\cite{semopt}. For Rust projects, we reproduce a Rust version under the same workflow and use the same project-level evaluation protocol. If a Rust project has no SemOpt optimization result above 5\%, the original workflow that first filters results above 5\% and then combines them cannot continue; in this case, we select the evaluable optimization result with the best performance as the final result.

\mainpoint{Codex Agent:}
The Codex Agent baseline uses OpenAI Codex~\cite{openai2026codex} to directly modify real projects, with three model configurations: gpt-5.3-codex-spark (Codex Spark)~\cite{openai2026gpt53codex}, gpt-5.4-mini (Codex 5.4 Mini)~\cite{openai2026gpt54mini}, and gpt-5.4 (Codex 5.4)~\cite{openai2026gpt54}. We use the parenthesized names as table abbreviations. All Codex results use the 30-minute version with high thinking mode. In the prompt, we provide the project repository, build script, test script, localized hotspot functions, performance optimization objective, and subsequent performance testing method.

\subsubsection{Models}\label{sec:real-world-models}

SemOpt and \toolname both use DeepSeek-V4-Pro~\cite{deepseek2026v4pro} as the optimizer model, keeping the default inference settings.

\subsubsection{Metrics}\label{sec:real-world-metrics}

For each optimization result, we apply it to the project, compile C/C++ and Rust code or install the Python package, and run project tests. Only passing results enter the performance statistics. For evaluable results, we run performance tests $6$ times for both unoptimized and optimized versions, discard the first run, and average the remaining $5$ runs. All project-level performance changes in Table~\ref{tab:real-world-performance-results} satisfy Welch's t-test at $p < 0.05$~\cite{welch1947generalization}. This repeated-measurement protocol is consistent with prior performance measurement studies~\cite{gao2024search, chen2016robust, liu2024minotaur, wen2025unveiling, fraile2025measuring}.

Different test cases may report performance metrics in different directions. If a larger value is better, let $x$ and $y$ denote results before and after optimization, respectively, and compute the improvement ratio as $\frac{y - x}{x}$. If a smaller value is better, the improvement ratio is computed as $\frac{x - y}{y}$. We use the following metrics to evaluate optimization effectiveness.
\begin{itemize}[leftmargin=*]
    \item \textbf{\#Pts.:} The number of data points used for performance statistics in each project. A data point corresponds to one test case or performance metric with independently computable performance improvement.
    \item \textbf{Max / Avg:} The maximum and average performance improvement ratio of a method over all data points in the project.
    \item \textbf{\# $\geq n\%$:} The number of data points whose performance improvement ratio reaches at least $n\%$. Because Table~\ref{tab:real-world-performance-results} separately reports total data points per project, the $\geq$5\% and $\geq$10\% columns report only threshold-reaching counts.
\end{itemize}

\subsubsection{Implementation Details}\label{sec:real-world-implementation-details}

For each project, \toolname identifies hotspot functions through profiling, generates candidate optimizations for them, and evaluates candidates under the same project testing and performance measurement protocol. For C/C++ projects, we use \code{perf}~\cite{perf} and \code{gcov}~\cite{gcov}. For Python projects, we use cProfile~\cite{python_profilers} and line\_profiler~\cite{lineprofiler}. For Rust projects, we use \code{perf} and locate hotspot functions using function symbols in the benchmark harness. We treat functions whose execution time exceeds $0.1\%$ of total runtime as hotspot functions.

For \toolname and SemOpt, we follow the SemOpt project-level setting~\cite{semopt}: a candidate is eligible for composition if it improves at least one data point by more than $5\%$ and degrades no other data point by more than $2\%$. For each hotspot function, we keep the eligible variant with the highest total improvement score, computed as the sum of improvement ratios over all data points, preserve the original function when none is eligible, and assemble the selected variants into the final optimized project version. For RQ4, each project-method pair contributes one final project-level result; Codex Agent's project patch is evaluated as that result. We then compute Max, Avg, $\geq$5\%, and $\geq$10\% over all data points. To reduce performance fluctuations, we avoid other resource-intensive programs during performance tests.

\subsection{RQ4: Real-world Optimization Results Compared with Baselines}\label{sec:real-world-rq4}

RQ4 evaluates project-level performance results of \toolname, SemOpt, and Codex Agent on real-world projects. Table~\ref{tab:real-world-performance-results} records each method's final result by project. All results use the same build, test, and performance measurement protocol.

\begin{table*}[t]
\centering
\caption{Project-level performance results in the real-world project optimization experiment. Under each method, the four columns report Max, Avg, \# $\geq$5\%, and \# $\geq$10\%, respectively.}
\label{tab:real-world-performance-results}
\scriptsize
\setlength{\tabcolsep}{2.6pt}
\renewcommand{\arraystretch}{1.05}
\setlength{\aboverulesep}{0.35ex}
\setlength{\belowrulesep}{0.35ex}
\setlength{\cmidrulesep}{0.15ex}
\begin{adjustbox}{max width=\textwidth,center}
\begin{tabular}{@{}llr*{5}{rrrr}@{}}
\toprule
\multirow{2}{*}{\textbf{Lang.}} & \multirow{2}{*}{\textbf{Repo}} & \multirow{2}{*}{\textbf{\#Pts.}}
 & \multicolumn{4}{c}{\textbf{SemOpt}}
 & \multicolumn{4}{c}{\textbf{Codex Spark}}
 & \multicolumn{4}{c}{\textbf{Codex 5.4 Mini}}
 & \multicolumn{4}{c}{\textbf{Codex 5.4}}
 & \multicolumn{4}{c}{\textbf{\toolname}} \\
\cmidrule(lr){4-7}\cmidrule(lr){8-11}\cmidrule(lr){12-15}\cmidrule(lr){16-19}\cmidrule(lr){20-23}
 & & & \textbf{Max} & \textbf{Avg} & \textbf{\#5} & \textbf{\#10}
   & \textbf{Max} & \textbf{Avg} & \textbf{\#5} & \textbf{\#10}
   & \textbf{Max} & \textbf{Avg} & \textbf{\#5} & \textbf{\#10}
   & \textbf{Max} & \textbf{Avg} & \textbf{\#5} & \textbf{\#10}
   & \textbf{Max} & \textbf{Avg} & \textbf{\#5} & \textbf{\#10} \\
\midrule
\multirow{5}{*}{C/C++}
  & RocksDB & 2 & 6.66\% & 2.38\% & \textbf{1} & 0 & 0.84\% & 0.19\% & 0 & 0 & -0.08\% & -0.40\% & 0 & 0 & 1.24\% & 0.83\% & 0 & 0 & \textbf{21.18\%} & \textbf{12.01\%} & \textbf{1} & \textbf{1} \\
  & Redis & 21 & 8.15\% & 1.20\% & 3 & 0 & 8.01\% & 0.25\% & 1 & 0 & 1.90\% & -0.20\% & 0 & 0 & 11.26\% & 1.99\% & 8 & 4 & \textbf{19.72\%} & \textbf{6.90\%} & \textbf{13} & \textbf{6} \\
  & gRPC & 16 & 37.25\% & 5.56\% & 8 & 7 & 6.39\% & 1.57\% & 1 & 0 & 22.18\% & 3.81\% & 7 & 3 & 11.01\% & 0.40\% & 2 & 1 & \textbf{67.55\%} & \textbf{16.69\%} & \textbf{10} & \textbf{9} \\
  & LevelDB & 22 & 160.36\% & \textbf{11.47\%} & \textbf{10} & 1 & 5.13\% & -0.31\% & 1 & 0 & 20.00\% & 0.91\% & 2 & 1 & 0.65\% & -0.89\% & 0 & 0 & \textbf{249.75\%} & 11.45\% & 6 & \textbf{4} \\
  & spdlog & 45 & 20.00\% & 3.33\% & 8 & 4 & 12.50\% & 2.06\% & \textbf{9} & 2 & 8.33\% & -0.46\% & 4 & 0 & 13.13\% & 0.12\% & 7 & 1 & \textbf{31.95\%} & \textbf{5.88\%} & 8 & \textbf{6} \\
\midrule
\multirow{5}{*}{Python}
  & Click & 11 & 410.24\% & 148.46\% & 10 & 10 & 20.56\% & 10.12\% & \textbf{11} & \textbf{11} & 5.28\% & 2.27\% & 2 & 0 & 46.51\% & 12.13\% & \textbf{11} & \textbf{11} & \textbf{717.42\%} & \textbf{258.17\%} & \textbf{11} & \textbf{11} \\
  & Flask & 8 & 159.35\% & 20.51\% & 1 & 1 & 334.14\% & 28.12\% & \textbf{4} & \textbf{4} & 0.85\% & 0.16\% & 0 & 0 & 3.52\% & 1.51\% & 0 & 0 & \textbf{346.23\%} & \textbf{44.55\%} & 1 & 1 \\
  & Jinja2 & 8 & 61.29\% & 32.20\% & 7 & 6 & 0.87\% & -0.68\% & 0 & 0 & 1.71\% & 0.18\% & 0 & 0 & 1.97\% & -0.15\% & 0 & 0 & \textbf{126.67\%} & \textbf{49.30\%} & \textbf{8} & \textbf{7} \\
  & Requests & 9 & 86.41\% & 42.46\% & \textbf{7} & 5 & 4.91\% & 2.92\% & 0 & 0 & 8.80\% & 4.36\% & \textbf{7} & 0 & 8.41\% & 5.05\% & \textbf{7} & 0 & \textbf{98.96\%} & \textbf{43.75\%} & \textbf{7} & \textbf{6} \\
  & Scrapy & 4 & 59.30\% & 53.44\% & \textbf{4} & \textbf{4} & 1.65\% & 0.28\% & 0 & 0 & 9.48\% & 1.73\% & 2 & 0 & 16.65\% & 6.74\% & 2 & 2 & \textbf{82.62\%} & \textbf{70.83\%} & \textbf{4} & \textbf{4} \\
\midrule
\multirow{5}{*}{Rust}
  & Rapier & 10 & 3.20\% & 0.71\% & 0 & 0 & NA & NA & NA & NA & -0.06\% & -8.34\% & 0 & 0 & -0.42\% & -1.47\% & 0 & 0 & \textbf{20.22\%} & \textbf{4.44\%} & \textbf{4} & \textbf{1} \\
  & image & 36 & 79.61\% & 16.87\% & 25 & 24 & 23.83\% & -0.14\% & 1 & 1 & 77.21\% & 0.87\% & 6 & 3 & 146.35\% & 13.86\% & 6 & 5 & \textbf{384.42\%} & \textbf{42.82\%} & \textbf{35} & \textbf{35} \\
  & regex & 16 & 29.89\% & 8.79\% & 6 & 5 & 74.43\% & 16.60\% & 8 & 7 & 12.70\% & 0.01\% & 1 & 1 & 18.28\% & 2.55\% & 3 & 2 & \textbf{83.46\%} & \textbf{23.17\%} & \textbf{11} & \textbf{11} \\
  & sqlparser & 8 & 8.49\% & 4.98\% & \textbf{5} & 0 & 3.36\% & 1.57\% & 0 & 0 & 4.30\% & 0.67\% & 0 & 0 & 9.50\% & 4.85\% & 1 & 0 & \textbf{23.81\%} & \textbf{5.08\%} & \textbf{5} & \textbf{1} \\
  & Tokio & 43 & 111.16\% & 4.57\% & 13 & 8 & 40.16\% & 3.46\% & 13 & 13 & 37.25\% & 3.21\% & 11 & 8 & 43.88\% & 1.84\% & 15 & 5 & \textbf{208.57\%} & \textbf{35.42\%} & \textbf{20} & \textbf{18} \\
\bottomrule
\end{tabular}
\end{adjustbox}
\end{table*}

\mainpoint{\toolname results.}
As shown in Table~\ref{tab:real-world-performance-results}, \toolname achieves the highest maximum performance improvement on all 15 projects and the highest average performance improvement on 14 projects. The maximum performance improvement ranges are 19.72\%--249.75\% on C/C++, 82.62\%--717.42\% on Python, and 20.22\%--384.42\% on Rust projects. The corresponding average performance improvement ranges are 5.88\%--16.69\%, 43.75\%--258.17\%, and 4.44\%--42.82\%. These results show that \toolname's gains do not depend on a single language or project and can produce measurable performance improvements in real projects from different ecosystems.

\mainpoint{Comparison with SemOpt.}
Compared with SemOpt, \toolname achieves a higher maximum performance improvement on every project, with relative improvements of 14.52\%--531.88\%. For average performance improvement, \toolname is lower by 0.02\% on LevelDB, while the other projects show relative improvements of 2.01\%--675.05\%. This indicates that \toolname can find higher-benefit optimization results on individual projects and improve overall average performance on most projects.

\mainpoint{Comparison with Codex Agent.}
Compared with Codex Agent, \toolname's advantage is more stable. Using the Codex version with the highest maximum performance improvement on each project as the comparison point, \toolname improves Max by 3.62\%--6329.95\% on comparable projects where that Codex version obtains a positive improvement. For Avg, \toolname also exceeds the best-performing Codex version on each project, with relative improvements of 4.74\%--27288.89\%. Codex can produce effective optimizations on some projects, and Codex Spark exceeds \toolname on Flask in both threshold-count metrics, \# $\geq$5\% and \# $\geq$10\%. However, although Codex Spark has a complete testing environment for compiling and running code, it has no usable performance result on Rapier because it fails at compilation, suggesting limitations in general code agents for project-level code optimization.

\begin{rqbox}
\textbf{Answer to RQ4.}
On the 15 real-world projects, \toolname ranks first in Max on all projects and in Avg on 14 projects, with relative improvements of 14.52\%--531.88\% in Max over SemOpt, a 0.02\% Avg decrease on LevelDB and 2.01\%--675.05\% Avg improvements on the other projects over SemOpt, and 3.62\%--6329.95\% in Max and 4.74\%--27288.89\% in Avg over the best Codex version on comparable projects.
\end{rqbox}

\subsection{RQ5: Strategy Source and Transfer Analysis}\label{sec:real-world-rq5}

RQ5 analyzes the source and cross-scenario example-transfer provenance of \toolname's effective optimization results. Table~\ref{tab:real-world-strategy-provenance} aggregates effective optimization results by language and traces their sources. Unlike RQ4, RQ5 counts all \toolname-generated effective optimization results before project-level final selection; multiple results from the same project, hotspot function, or selected rule are counted separately when they are distinct evaluated optimization outputs. \textbf{Effective} denotes optimization results that pass project tests and obtain a statistically significant positive performance improvement ($p < 0.05$) under the measurement protocol in Section~\ref{sec:real-world-metrics}. \textbf{Pre-transfer} denotes effective results whose selected rules come from candidates already available in the target scenario before cross-scenario construction. \textbf{Post-transfer} denotes effective results whose selected rules come from target-scenario candidates constructed using cross-scenario example transfer or descriptions from other scenarios. \textbf{Has doc info} denotes effective results whose sources contain external documentation information. \textbf{x86\_64} reports source architecture provenance from the selected strategy cluster.

\begin{table}[t]
\centering
\caption{Source and cross-scenario example-transfer provenance of effective optimization results. Percentages use the Effective count in the same row as the denominator.}
\label{tab:real-world-strategy-provenance}
\scriptsize
\setlength{\tabcolsep}{1.2pt}
\renewcommand{\arraystretch}{1.08}
\begin{adjustbox}{max width=\columnwidth}
\begin{tabular}{@{}lrrrrr@{}}
\toprule
\textbf{Lang.} & \textbf{Eff.} & \textbf{Pre} & \textbf{Post} & \textbf{Doc} & \textbf{x86\_64} \\
\midrule
C/C++ & 1404 & 1178 (83.90\%) & 226 (16.10\%) & 642 (45.73\%) & 642 (45.73\%) \\
Python & 1998 & 489 (24.47\%) & 1509 (75.53\%) & 161 (8.06\%) & 0 (0.00\%) \\
Rust & 429 & 39 (9.09\%) & 390 (90.91\%) & 37 (8.62\%) & 2 (0.47\%) \\
\midrule
Total & 3831 & 1706 (44.53\%) & 2125 (55.47\%) & 840 (21.93\%) & 644 (16.81\%) \\
\bottomrule
\end{tabular}
\end{adjustbox}
\end{table}

\noindent\textbf{Documentation information} provides an important supplement for effective optimization results. Overall, 840 of 3831 effective results contain documentation information, accounting for 21.93\%. C/C++ has the largest proportion, at 45.73\%. Because the current external documentation information only covers C/C++-side documentation, the lower Has doc info proportions for Python and Rust are expected. This is notable because documentation inputs are much smaller than commit-derived inputs, yet they account for nearly half of effective C/C++ results. Even without Python or Rust documentation inputs, documentation information appears in 8.06\% and 8.62\% of their effective results, suggesting that it may also contribute through cross-scenario target-candidate construction.

\noindent\textbf{Post-transfer target-scenario candidates.} Cross-scenario example transfer also contributes many effective optimization results through target-candidate construction. Overall, 55.47\% of effective results come from post-transfer target-scenario candidates, exceeding the 44.53\% from pre-transfer candidates. This contribution is particularly clear in Python and Rust, where post-transfer rules account for 75.53\% and 90.91\% of effective results, respectively. This shows that \toolname does not depend only on candidates already available in the target scenario, but also benefits from additional optimization knowledge introduced from other scenarios.

\begin{rqbox}
\textbf{Answer to RQ5.}
Among \toolname's effective optimization results, documentation information appears in 21.93\% overall and 45.73\% for C/C++, while post-transfer target-scenario candidates account for 55.47\%.
\end{rqbox}

\section{Threats to Validity}\label{sec:discussion-threats}

The main threats come from data leakage, benchmark and model choices, construct metrics, and implementation. We exclude each task's \emph{exact commit} or \emph{exact code source item} and compare methods with the same LLM when possible, but LLM training data cannot be fully audited. EM and SemEqv measure reproduction of historical edits, while Max and Avg depend on available project workloads. C/C++ and Python reuse SemOpt's open-source code, benchmarks, and paper settings, while Rust is reimplemented following the SemOpt workflow. These choices improve comparability but bind the conclusions to this benchmark family and evaluation protocol. LLM nondeterminism, rule-generation failures, data-processing bugs, and scripted evaluation errors may also affect the results.

\section{Related Work}\label{sec:related-work}

Code optimization has long used rule-based, program-analysis, and search techniques, including compiler optimizations, memoization, misconfiguration repair, genetic optimization, CLion, and Clang-Tidy~\cite{mckeeman1965peephole,cocke1970global,della2015performance,krishna2020cadet,giavrimis2021genetic,clion,clangtidy}. These methods handle specific inefficiency patterns, but depend on expert-defined rules and are difficult to extend across languages and scenarios. Dedicated performance optimization models, such as VQ-VAE, Supersonic, and DeepDev-PERF~\cite{chen2022learning,chen2024supersonic,garg2022deepdev}, often underperform LLM-based methods~\cite{garg2023rapgen}.

Recent studies have explored LLM-based code optimization. PIE constructs an efficiency benchmark and finds that RAG outperforms direct prompting~\cite{shypula2023learning}; Gao et al. further study how example selection and ordering affect RAG~\cite{gao2023makes}. We use RAG as a baseline to test whether raw retrieval can replace explicit strategy construction. RAPGen targets C\# API-replacement patches and mainly supports a single optimization strategy~\cite{garg2023rapgen}. SemOpt extracts strategies from optimization commits and generates associated rules~\cite{semopt}; in contrast, \toolname{} constructs strategies from heterogeneous data sources and transfers them across language and architecture scenarios. Cross-language code intelligence studies high-resource to low-resource code generation and multilingual code capabilities~\cite{cassano2024knowledge,giagnorio2025enhancing,joel2025survey}; \toolname{} instead transfers optimization strategies and regenerates target-scenario localization rules. Other systems are complementary: SBLLM searches candidate versions~\cite{gao2024search}, SWIFTCODER fine-tunes for efficient code~\cite{huang2024swiftcoder}, PerfCodeGen uses execution feedback~\cite{peng2025perfcodegen}, PEACE studies project-level Python optimization through dependency-aware hybrid code editing~\cite{ren2025peace}, and POLO uses profiling, project context, and agent feedback for project-level C and C++ optimization~\cite{bai2025polo}. These systems improve search, training, feedback, or project-level editing, whereas \toolname{} focuses on strategy acquisition, abstraction, cross-scenario transfer, and rule-based localization.

Recent studies have also used LLMs to synthesize static analysis checkers, including KNighter, AutoChecker, and MoCQ~\cite{yang2025knighter,liu2024write,li2025automated}. These studies show that high-precision checker synthesis still faces semantic boundaries, false positives, and false negatives. In \toolname{}, Semgrep rules are only optimization-opportunity locators, not patch templates. Examples validate rule behavior, while final correctness is handled by LLM optimization, compilation and testing, and performance validation. Therefore, the rule stage can place more emphasis on recall.

\section{Conclusion}\label{sec:conclusion}

This paper presents \toolname, an LLM-based code optimization framework that addresses two limitations of strategy-guided optimization: reliance on historical commits and source-scenario-specific strategy formalization.
\toolname represents heterogeneous knowledge-source items as evidence objects, balances them through self-balanced weighted clustering, constructs target-scenario examples when needed, and generates validated static analysis rules to guide LLM optimization.
Experiments on 351 historical optimization tasks and 15 real-world projects show that \toolname improves optimization reproduction and project-level performance over SemOpt and Codex.
Overall, the results indicate that integrating multiple knowledge sources across scenarios can effectively expand strategy coverage for LLM-based code optimization.

\section*{Data Availability}

Artifacts of this paper, including the implementation of \toolname, all baseline implementations, and evaluation results, are available at \url{https://figshare.com/s/f4499b791389351c7217}.

\balance
\bibliographystyle{bibtex/IEEEtran}
\bibliography{references}

\end{document}